\documentclass[aps,prl,twocolumn,showpacs]{revtex4}
\usepackage{graphicx}

\begin{document}

\title{Controlled dephasing of a quantum dot in the Kondo regime}
\author{M.~Avinun-Kalish, M.~Heiblum, A.~Silva, D.~Mahalu, and V.~Umansky}
\address{Braun Center for Submicron Research, Department of Condensed
 Matter Physics, Weizmann Institute of Science, Rehovot 76100, Israel}
\date{\today}

\begin{abstract}
Kondo correlation in a spin polarized quantum dot (QD) results
from the dynamical formation of a spin singlet between the dot's
net spin and a Kondo cloud of electrons in the leads, leading to
enhanced coherent transport through the QD.  We demonstrate here
significant dephasing of such transport by coupling the QD and its
leads to potential fluctuations in a near by 'potential detector'.
The qualitative dephasing is similar to that of a QD in the
Coulomb Blockade regime in spite of the fact that the mechanism of
transport is quite different. A much stronger than expected
suppression of coherent transport is measured, suggesting that
dephasing is induced mostly in the 'Kondo cloud' of electrons
within the leads and not in the QD.
\end{abstract}


\maketitle

Experiments involving \textit{controlled dephasing} of coherent
systems are excellent means for probing the nature of phase
coherent transport. Such experiments were recently performed in
mesoscopic structures based on \textit{quantum dots} (QD) in the
\textit{Coulomb Blockade} (CB) regime~\cite{r1,r2,Sprinzak}.
Coherence of the QD was monitored by embedding it in a two path
\textit{Aharonov-Bohm} (AB) interferometer. Dephasing was induced
by a capacitively coupled \textit{quantum point contact} (QPC)
which serves as a \textit{which path} detector (a detector
sensitive to the electron trajectory). We now perform  a
controlled dephasing experiment of a QD in the Kondo correlated
regime, where transport is ideally expected to be fully coherent.

Kondo correlation takes place when a QD, containing an odd number
of electrons and a net spin, has strong coupling to its leads. The
strong coupling allows spilling of the electronic wave-function of
the highest occupied energy level in the dot to the leads. This
enables the formation of a dynamical many-body spin singlet,
spanning across a \textit{Kondo cloud}~\cite{Aleiner}, and a
resonance in the QD density of states at the Fermi energy (as a
consequence of a coherent superposition of spin flip co-tunnelling
events)~\cite{Anderson}. The characteristic energy scale of this
dynamical singlet is termed the \textit{Kondo temperature},
$T_{k}$, and the typical size of the Kondo cloud is $\frac{\hbar
v_{F}}{k_{B}T_{k}}$, with $v_{F}$ the Fermi velocity, h Plank's
constant, and $k_{B}$ the Boltzman constant. This enhances the
conductance, $G_{QD}$, and for an electron temperature $T \ll
T_{k}$ (the \textit{unitary limit}),
$G_{QD}=2e^{2}/h$~\cite{DGGkondo,Wiel,Cronenwett}. The application
of either a small \textit{source-drain} bias $V_{sd}$, a finite
magnetic field, or an increased temperature, will all quench the
Kondo correlation. Since correlation is only among the spins of
the electrons, one may ask whether dephasing can be induced via
nearby potential fluctuations that couple only to the charge. It
was predicted~\cite{Glazman,Silva} that under weak interaction
between a 'Kondo dot' and an adjacent biased QPC the \textit{Kondo
valley} enhanced conductance will be suppressed.

\begin{figure*}
\includegraphics[width=7.0in]{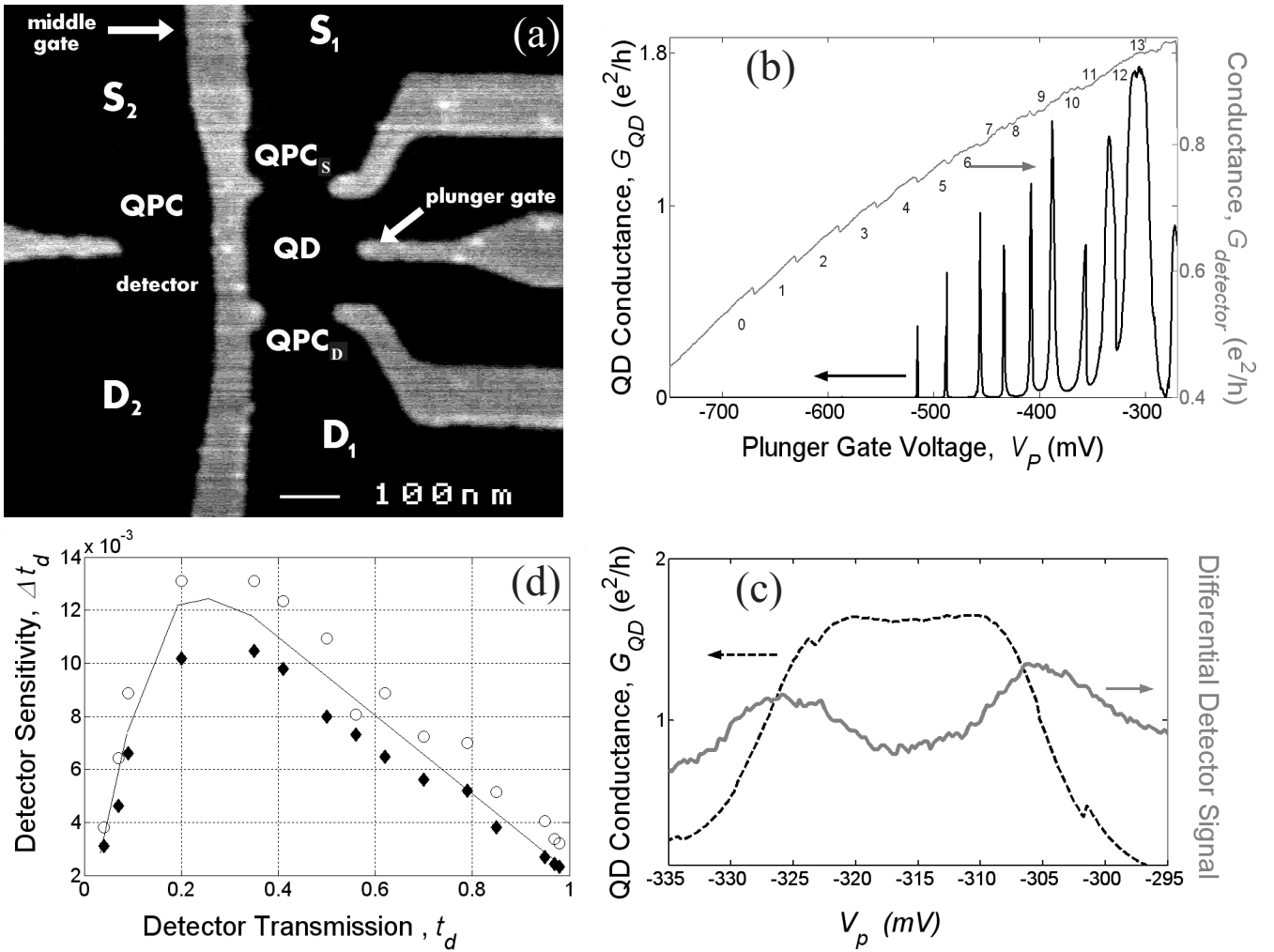}
\caption{ (\textbf{a}) Top view SEM image of the structure.  The
gray areas are metallic gates deposited on the surface of a
GaAs-AlGaAs heterojunction embedding the 2DEG. (\textbf{b})
Conductance of the QPC detector (top gray trace) and conductance
of the QD (the lower black trace) as a function of the plunger
gate voltage, $V_{p}$. The numbers 0-13 indicate the number of
electrons in the QD. (\textbf{c}) The black trace presents a Kondo
peak, namely conductance $G_{QD}$ as a function of $V_{p}$, while
the gray trace is the detector differential signal, namely the
derivative of the current through the QPC with respect to the
modulation of the QD plunger gate. Here $V_{d}=1.5meV$ and
$t_{d}=0.3$. (\textbf{d}) Detector sensitivity $\Delta t_{d}$ as a
function of $t_{d}$ (Here $T_{k1} \sim 41\mu eV$). Measurements
are made on both sides of the Kondo resonance: the circles between
12 and 13 electrons while the diamonds between 13 and 14
electrons.}\label{Fig1}
\end{figure*}

One can view the QPC as a detector or alternatively as a noise
source~\cite{r1,r2,Sprinzak}. It serves as a detector since its
conductance is affected by long living trajectories in the Kondo
system, and this detection prevents coherent superposition. On the
other hand, the QPC serves as a noise source via shot noise
fluctuations in its current~\cite{reznikov}: $S\propto
I_{DC}t_{d}(1-t_{d})$, with $t_{d}$ the transmission of the QPC
and $I_{DC}$ the impinging current. The transmission resonance of
a QD in the CB regime broadens proportionally to the dephasing
rate induced by the QPC~\cite{Sprinzak}. By contrast, theory
predicts~\cite{Silva} that the width of the Kondo resonance,
$k_{B}T_{k}$, will remain unaffected by the interaction with the
QPC. Here we report the results of such a study, in which we
observe quenching of the otherwise enhanced \textit{Kondo valley}
as a result of interaction with a nearby QPC detector. We find a
much stronger effect than anticipated~\cite{Silva}, pointing at
the role of the Kondo cloud. Nevertheless, our experimental setup
prevents us from being able to prove that $T_{k}$ remains
unaffected by the dephasing process.

Buks \textit{et al.} had already demonstrated~\cite{r1} the
suppression of coherent transmission of a QD in the CB regime via
\textit{path detection} by a biased QPC. This was done by
embedding the QD in an Aharonov-Bohm interferometer and monitoring
the visibility of the interference pattern $\nu _{d} \sim
1-\gamma$ (with $\gamma$ the \textit{suppression strength}), as a
function of $t_{d}$ and the applied bias $V_{d}$ across the QPC.
One obtains~\cite{r1,r2} a dependence $\gamma = \frac{1}{8\pi}\
\frac{eV_{d}}{\Gamma}\ \frac{(\Delta t_{d})^2}{t_{d}(1 - t_{d})}\
\equiv \frac{\Gamma_{d}(t_{d},\Delta t_{d},V_{sd})}{\Gamma}$, with
$\Gamma$ the unperturbed single particle level width and $\Delta
t_{d}$ the change of the detector's transmission $t_{d}$ due to an
added electron in the QD.  Note that the suppression of the
visibility is accompanied by broadening of the resonance peak
$\Gamma\ \rightarrow \Gamma + \Gamma _{d}$ . In contrary, if we
assume only QPC-QD interaction in the Kondo regime, the
suppression of the coherent transport has an almost identical
form, but with $k_{B}T_{k}$ replacing $\Gamma$: $\gamma =
\frac{1}{8\pi}\ \frac{eV_{d}}{k_{B}T_{k}}\ \frac{(\Delta
t_{d})^2}{t_{d}(1 - t_{d})}\ $. However, unlike the CB case, the
width of the Kondo resonance is expected to remain (to first
approximation) $k_{B}T_{k}$~\cite{Silva}.

The experimental configuration, shown in Fig. 1, consists of two
neighboring systems: a QD on the right hand side and a QPC on the
left, each with its separate current path.  Confinement in the two
dimensional electron gas (2DEG; 53 nm below the surface; areal
electron density $3.3\times 10^{11} cm^{-2}$; mobility $\mu =
1.6\times 10^{6} cm^{2}/V s$ at 4.2K) is provided by negatively
biased metallic gates deposited on the surface of the structure.
The QD is well coupled to the leads (via its confining QPCs),
hence enabling the formation of the Kondo cloud. The separate QPC
on the left - serves as a sensitive potential detector - partly
reflecting the incident current with conductance $G_{detector}=
t_{d} \frac{2e^{2}}{h}\ $. An electron added to the QD, responding
to an increased voltage of the plunger gate, screens the gate
potential and changes the transmission of the QPC by $\Delta
t_{d}$ (acting as a 'far away gate'). This detection method can
also be used to count electrons, as demonstrated in Fig. 1(b),
where the detector signal is shown at the top trace and the
conductance of the QD at the lower one. The \textit{saw-tooth}
pattern of conductance of the QPC, as a function of plunger gate
voltage, reflects the potential evolution in the QD. Note that the
saw-tooth behavior persists even though the conductance peaks of
the QD cannot be resolved anymore, with a clearly observed
\textit{first electron} in the dot. The strong enhancement in the
conductance valley for 13 electrons in the QD indicates that the
QD is in the Kondo correlated regime - as was indeed verified by
raising the temperature, applying a DC voltage, or increasing the
magnetic field.

\begin{figure}
\includegraphics[width=3.5in]{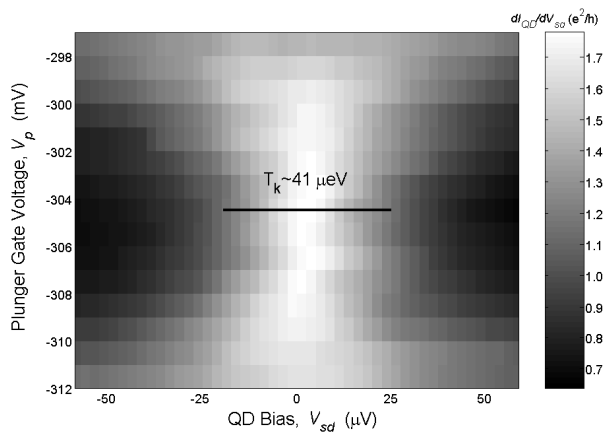}
\caption{Differential conductance of the QD as a function of
plunger gate voltage and applied bias.  The ZBA's FWHM, measured
at the narrowest neck of the conductance peak, yields $T_{k}$.
 }  \label{Fig2}
\end{figure}

We first characterize the detector by measuring its sensitivity
$\Delta t_{d}$ as a function of its transmission $t_{d}$. Since
the QD is relatively open in the Kondo regime the
\textit{saw-tooth} potential is smoothed out and $\Delta t_{d}$ is
very difficult to determine accurately (see Fig. 1(b) near the
13th electron). Hence, we perform a differential measurement
$\frac{dI_{QPC}}{dV_{p}}$, by modulating the QD's plunger gate
voltage, the results are shown in Fig 1(c). The two peaks are
indicative of two electrons added to the dot, and the integral
under each peak provides the detector's sensitivity, $\Delta
t_{d}$. It peaks near $t_{d} \sim 0.3$ and diminishes, as
expected, at the conductance plateaus ($t_{d}=0$ and $t_{d}=1$,
Fig. 1(d)). The Kondo temperature was estimated from the full
width at half maximum (FWHM) of the \textit{Zero Bias Anomaly}
(ZBA), namely, the width of the differential conductance peak as a
function of the voltage across the QD, $V_{sd}$.  The width was
determined from the 3D plot of the differential conductance as a
function of both $V_{sd}$ and the \textit{plunger gate} voltage,
$V_{p}$ (Fig. 2). We find that the width, $\sim
k_{B}T_{k}$~\cite{ZBA}, is temperature independent up to $T \sim$
50mK ($k_{B}T \sim 4 \mu eV$) with $T_{k1} \sim 41\pm 3 \mu eV$.
Above 50mK the temperature dominates the FWHM, which increases
linearly with temperature with a slope of $\sim5.4k_{B}$; in
agreement with previous reports~\cite{Wiel,Cronenwett}. For a
lower Kondo temperature (set by reducing the coupling to the
leads), $T_{k2} \sim 12\pm 5 \mu eV$, the FWHM of the ZBA is
already in the temperature dependent regime at our base
temperature (35mK, $k_{B}T \sim 3 \mu eV$). Then $T_{k}$ is
estimated from the slope of the FWHM vs.
temperature~\cite{comment_detTk}.

\begin{figure}
\includegraphics[width=3.5in]{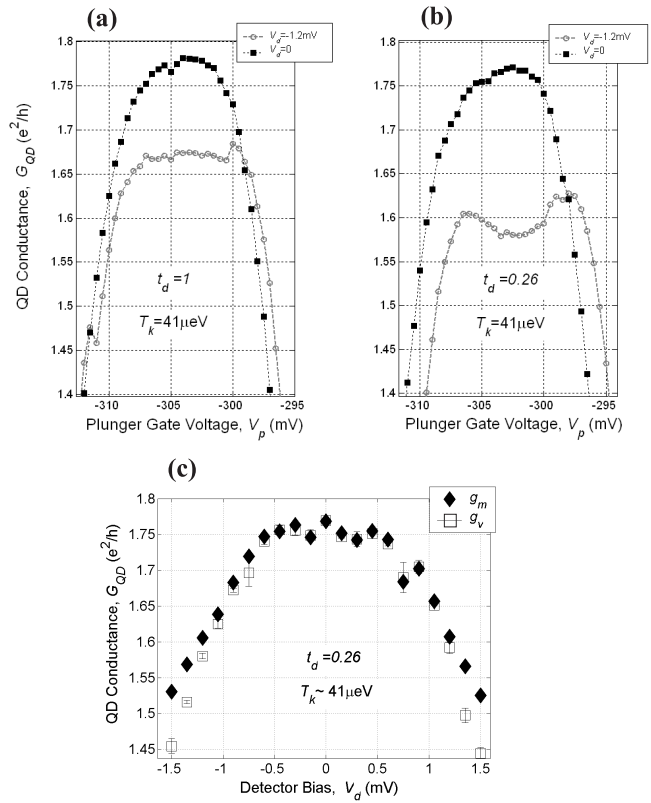}
\caption{(\textbf{a}) Conductance $G_{QD}$ as a function of
$V_{p}$ for $V_{d}=0$ and $1.2mV$, when the detector is tuned to
$t_{d}=1$. (\textbf{b}) The same data as in (a) but for
$t_{d}=0.26$.  The applied bias on the detector SD serves as a
gate, which shifts the conductance picture, and breaks the
symmetry of the two QPCs forming the QD. This results in a uniform
suppression of conductance at the peaks. However the dip, which is
developed in the Kondo valley for $t_{d}\neq 1$, is a consequence
of the dephasing interaction with the detector. (\textbf{c})
$g_{m}$ (full diamonds) and $g_{v}$ (empty squares) as a function
of $V_{d}$, for $t_{d}=0.26$.
 }  \label{Fig3}
\end{figure}

An unintentional electrostatic coupling between the QPC and the QD
is always present due to their proximity.  For example, changing
$t_{d}$ of the QPC via its gate voltage inadvertently alters the
coupling of the QD to the leads. Moreover, applying $V_{d}$ across
the QPC raises the potential on one side of the QPC and affects
the delicate symmetry of the two QPCs that form the QD. Hence
$\Gamma_{S} \neq \Gamma_{D}$ and the transmission of the QD
decreases. These \textit{extrinsic} effects, which might be
interpreted erroneously as dephasing, must be taken into account.
Therefore, for each setting of $t_{d}$, $QPC_{S}$ and $QPC_{D}$ of
the QD (see Fig.1(a)) were readjusted in order to keep $T_{k}$
constant and maximize the linear conductance of the valley ($\sim
1.8 e^{2}/h$) at $V_{sd}=0$. In order to correct the undesirable
\textit{gating} effect which is due to the bias of the QPC
detector, the conductance of the QD was renormalized: either by
the value of the conductance at $t_{d}=1$, where there are no shot
noise fluctuations in the QPC and therefore no dephasing is
expected (see Fig. 3(a)); or by the conductance peaks on both
sides of the Kondo valley, where the dephasing rate is expected to
be negligibly small but symmetry of the QD is important in
determining the heights of the peaks. The latter procedure is
demonstrated in Fig. 3(b) and the data are summarized in Fig.
3(c). There we plot the valley conductance $g_{v}$ and the peaks
conductance $g_{m}$ as a function of $V_{d}$ across the QPC. The
normalized suppression strength is $\gamma=1-\frac{g_{v}}{g_{m}}\
\frac{g_{m0}}{g_{v0}}\ $, where $g_{v0}$ and $g_{m0}$ are the
corresponding QD conductances for $V_{d}=0$.

The suppression strength, $\gamma$ , for two different Kondo
temperatures exhibits a double-hump behavior as a function of
$t_{d}$ (Figs. 4(a) and 4(b)).  The suppression strength is found
to be inversely proportional to $T_{k}$~\cite{r13}, namely,
$\gamma_{1}T_{k1} \cong \gamma_{2}T_{k2} $ for each value of the
detector's parameters ($V_{d}, t_{d}$). A comparison with the
theoretical prediction of the suppression strength~\cite{Silva},
substituting the measured $\Delta t_{d}$ (found in Fig. 2),
highlights three distinct discrepancies. The first is a large
quantitative disagreement, namely, our measured suppression
strength is some 30 times larger. The second is the quadratic bias
dependence (data not shown here), while the theory predicts a
linear dependence. The third, seen in Fig. 4(c), is a deviation
from the otherwise qualitative agreement near the second hump,
namely near $t_{d}\sim 0.7$. We find a strong dephasing peak which
the theory doesn't predict.

\begin{figure}
\includegraphics[width=3.5in]{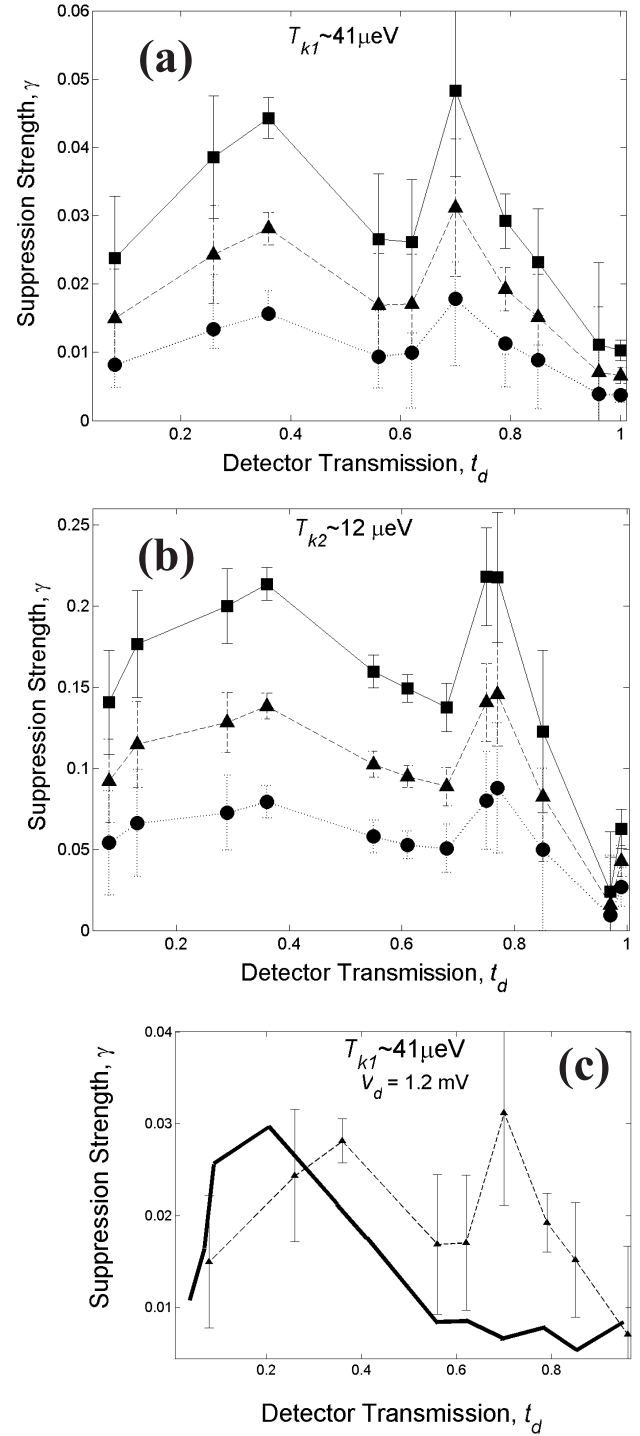}
\caption{The suppression strength ($\gamma$) as a function of
$t_{d}$ for a few values of $V_{d}$ (squares $V_{d}=1.5mV$;
triangles $V_{d}=1.2mV$; circles $V_{d}=0.9mV$).  Two sets are
presented: (\textbf{a}) $T_{k1} \sim 41\pm 3\mu eV$, (\textbf{b})
$T_{k1} \sim 12\pm 5\mu eV$. (The error bars are determined by the
least-square fit of a quadratic function of $V_{d}$ for fixed
$T_{k}$ and $t_{d}$).(\textbf{c}) An example of the measured and
the predicted effect multiplied by 30 (diamonds) for $V_{d}=1.2mV$
and $T_{k1} \sim 41\pm 3\mu eV$ (using the measured values of
$\Delta t_{d}$ and $t_{d}$). }  \label{Fig4}
\end{figure}

 Though we do not have definite explanations to these
discrepancies we wish to suggest the following possibilities. The
theory in Ref. 10 follows similar arguments to those for a QD in
the CB regime~\cite{r1,r2}, assuming only QPC-QD interaction. In
the CB regime it is a justified assumption, since the electron is
confined to the QD with a dwelling time $\sim1/\Gamma$. However,
in the Kondo regime, an electron transmitted from source to drain
dwells for a long time (approximately $\hbar/ k_{B}T_{k}$) inside
the 'Kondo cloud', comprising the QD and part of the leads. Since
the actual phase space for scattering in the 2D leads (having
continuous density of states) is much greater, the dephasing of
the leads electrons inside the Kondo cloud is dominant. Employing
the 'Fermi golden rule', one gets a similar dependence as for e-e
scattering, leading to $\gamma\sim
\frac{(eV_{d})^{2}}{k_{B}T_{k}}$ instead of $\gamma\sim eV_{d}$ is
predicted in Ref 10. This modified expression results in a
reasonable quantitative agreement and at the same time explains
the quadratic dependence we found on $V_{d}$. These results
demonstrate, even though not directly, the existence of the Kondo
cloud. The qualitative discrepancy near $t_{d}\sim 0.7$ might be
related to the so-called "0.7 anomaly"~\cite{r14}, where the shot
noise is expected to be suppressed~\cite{r15} relative to the
noise $S\propto t_{d}(1-t_{d})$ of the QPC. Our QPC detector does
not show a distinct 0.7 plateau in its conductance plot vs. gate
voltage at $T= 35mK$, but only a slight change in the curvature.
However, it does show a small ZBA peak in its differential
conductance in the vicinity of $t_{d}\sim 0.7$~\cite{CM07},
suggesting a deviation from the ubiquitous behavior of the QPC.

An important prediction is that $T_{k}$ remains fixed independent
of the dephasing rate, even though the many body transmission is
strongly affected by the dephasing rate.  This is in contrast with
the dephasing process in a Coulomb Blockaded QD, where the level
width $\Gamma$ is directly related to the dephasing rate.
Unfortunately, our experimental setup did not allow an exact
determination of $T_{k}$ as a function of $V_{d}$ (any better than
10 \% accuracy). Hence we can only claim that $T_{k}$ didn't
change by more than 10 \%, prohibiting an exact verification of
this point.

In conclusion, we have studied the response of a QD in the Kondo
correlated regime to an interaction with charge fluctuations in a
nearby, capacitively coupled, QPC detector. Despite the
differences between the transport in a QD in the Kondo correlated
regime and a QD in the CB regime, we find similar characteristic
dephasing behavior in both. We show that the suppression strength
is inversely proportional to the Kondo temperature. Moreover, we
can explain our results only by assuming on the existence of an
extended Kondo cloud outside the QD.

\textbf{Acknowledgments} We thank Ji~Yang, Y.~Oreg and S.~Levit
for fruitful discussions.  The research was partly supported by
the Israeli Ministry of Science, the MINERVA foundation, The
German Israeli Project Cooperation (DIP), the German Israeli
Foundation (JIF), and the EU QUACS network.


\end{document}